# Observation of the frozen charge of a Kondo resonance

M.M. Desjardins[1], J.J. Viennot[2], M.C Dartiailh[1], L.E. Bruhat[1], M.R. Delbecq[3,†], M. Lee[4], M.-S. Choi[5], A. Cottet[1] and T. Kontos[1,*]

[1] *Laboratoire Pierre Aigrain, Ecole Normale Supérieure-PSL Research University, CNRS, Université Pierre et Marie Curie-Sorbonne Universités, Université Paris Diderot-Sorbonne Paris Cité, 24 rue Lhomond, 75231 Paris Cedex 05, France*

[2] *JILA and Department of Physics, University of Colorado, Boulder, Colorado, 80309, USA*

[3] *Center for Emergent Matter Science, RIKEN, 147 Main Bldg., 2-1 Hirosawa, Wako-shi, Saitama, 351-0198 Japan*

[4] *Department of Applied Physics, College of Applied Science, Kyung Hee University, 1732 Deogyeong-daero, Giheung-gu, Yongin-si, Gyeonggi-do, 446-701, Republic of Korea*

[5] *Department of Physics, Korea University, 145 Anam-ro, Seoul 02841, South Korea*

\*To whom correspondence should be addressed: kontos@lpa.ens.fr

†Now at: JEIP, USR 3573 CNRS, Collège de France, PSL Research University, 11, place Marcelin Berthelot, 75231 Paris Cedex 05, France

**The ability to control electronic states at the nanoscale has contributed to our modern understanding of condensed matter. In particular, quantum dot circuits represent model systems for the study of strong electronic correlations, epitomized by the Kondo effect[1,2,3]. Here, we show that circuit Quantum Electrodynamics architectures can be used to study the internal degrees of freedom of such a many-body phenomenon. We couple a quantum dot to a high finesse microwave cavity to measure with an unprecedented sensitivity the dot electronic compressibility i.e. the ability of the dot to accommodate charges. Because it corresponds solely to the charge response of the electronic system, this quantity is not equivalent to the conductance which involves in general other degrees of freedom such as spin. By performing dual conductance/compressibility measurements in the Kondo regime,**

**we uncover directly the charge dynamics of this peculiar mechanism of electron transfer. Strikingly, the Kondo resonance, visible in transport measurements, is 'transparent' to microwave photons trapped in the high finesse cavity. This reveals that, in such a many body resonance, finite conduction is achieved from a charge frozen by Coulomb interaction. This previously elusive freezing of charge dynamics[4,5,6] is in stark contrast with the physics of a free electron gas. Our setup highlights the power of circuit quantum electrodynamics architectures to study condensed matter problems. The tools of cavity quantum electrodynamics could be used in other types of mesoscopic circuits with many-body correlations[7,8] and bring a promising platform to perform quantum simulation of fermion-boson problems.**

In a free electron gas, electrical conduction is carried by mobile charges. Its compressibility $\chi = \frac{\partial N}{\partial \mu}$ with $N$ the number of electrons and $\mu$ the chemical potential is simply the density of states at the Fermi energy. It is therefore directly linked to the finite conductivity of the system. This explains for example why both the compressibility and the conductivity provide essentially the same piece of information for alkali metals. But what happens in the case of a strongly correlated electronic gas? A paroxysmal situation is that of an electron localized on a single site with strong Coulomb repulsion, coupled to a continuum of electronic states[1-4,7-9]. Through its link to the Kondo problem, such a configuration, besides its apparent simplicity, is relevant for understanding different types of strongly correlated gases, ranging from heavy fermions to high Tc superconductors[4,9], and therefore a priori relevant for many condensed matter problems.

A single localized level is expected to have a much smaller electronic compressibility than a piece of metal, since its density of states at the Fermi energy is dramatically reduced. How one could measure the tiny compressibility of a single localized state? Such a measurement requires first to isolate in a controlled manner a single electron, which can be conveniently done using a quantum dot (QD) circuit, but also to measure its tiny effective capacitance, which is equivalent to the compressibility of an electron gas. Although this can be done using low frequency as well as microwave techniques[10-13], it has been shown recently that this could be achieved alternatively with an unprecedented sensitivity using a circuit Quantum Electrodynamics architecture[14]. Importantly, since the first compressibility measurements in quantum dots[10], correlation effects have been sought for. Our work is the first where correlations effects are directly (and qualitatively) visible in the compressibility of a quantum dot circuit[5,6]. For that purpose, we use the prototype example of the Kondo regime. The principle of our measurement architecture is shown in figure 1a: the finite compressibility $\chi$ shifts the frequency of the microwave resonator (as shown in figure 1b), used here as a non-invasive probe (see Methods). This frequency shift, read-out from the phase of the microwave signal, is only sensitive to variations of the dot charge, in contrast with the conductance for which all degrees of freedom can contribute (charge and spin). The linewidth of the cavity and the electron-photon coupling strength set the limit to the smallest detectable $\chi$.

The experimental setup is shown in figure 1c. A single quantum dot circuit made out of a single wall carbon nanotube is embedded in a coplanar wave guide cavity and coupled capacitively[15] to it (see Methods). We measure simultaneously the DC current flowing through the quantum dot and the phase and amplitude of the transmitted microwave signal at the cavity frequency ($f_{cav}$=6.67129 GHz). Such a setup allows us to characterize accurately the electron-photon interaction- which is essential for performing a compressibility measurement.

Carbon nanotube QDs can be tuned from the deep Coulomb blockade regime to the Kondo regime[16] simply by changing the voltage $V_g$ applied to an electrostatic gate, as the one coloured in green in figure 1c. For low gate voltages, we observe standard Coulomb diamonds in the $V_g$-$V_{sd}$ plane, where $V_{sd}$ is the source(S)-drain(D) bias. Figure 2a and 2b display the characteristic periodic patterns of the conductance and the microwave phase in this regime at the base temperature of our experiment $T$=255mK. The conductance resonances delimiting the Coulomb diamonds appear simultaneously as peaks of about 3° in the phase signal. This indicates a finite compressibility of the QD electron gas for these resonances. In contrast, in a Coulomb valley, the charge dynamics in the dot is frozen, which leads to the absence of compressibility as shown in figure 2d. The finite compressibility and the peaks in the conductance are perfectly correlated (see figure 2c), which is reminiscent of a weakly correlated electron gas although interactions are present manifested by Coulomb blockade.

The nature and the strength of the electron-photon coupling in our device is calibrated using a well-known situation, the Coulomb blockade in the linear regime (bias $eV_{sd}<k_BT$, figure 2c). In that case, electron transport only occurs when the electron gas in the QD has a finite density of states at the Fermi energy, which also corresponds to a finite compressibility[17-19]. Because the finite compressibility is linked to back and forth tunnelling of electrons between the QD and the leads, it creates a dipole which couples to the cavity (see figure 2e, top panel). The compressibility can therefore be read out through a shift of the resonance frequency of the cavity. This yields the corresponding phase shift $\Delta\varphi = 2g^2\hbar\chi/\kappa$ for the transmitted microwave signal, where $\kappa$ and $g$ correspond respectively to the linewidth of the cavity and the electron-photon coupling strength (see Methods). The compressibility of the quantum dot depends on the linewidth of a Coulomb peak. Using a well-established theory (see Methods), we find for all the Coulomb peaks studied $g \sim 2\pi$ x (65MHz+/-15MHz) [20]. In that case, the compressibility $\frac{\partial N}{\partial \varepsilon_d}$ can also be viewed as the zero-frequency charge susceptibility, which stems from the retarded correlator $\chi(t) = -i\theta(t)\langle[\hat{n}(t),\hat{n}(0)]\rangle$, where $\hat{n}(t)$ is the electron number operator of the dot and $\theta(t)$ is a step function (see Methods). As a conclusion, in the Coulomb blockade regime, both finite conductance and compressibility only arise from the ability of the mobile charges to tunnel in or out of the dot. Importantly, our cQED architecture resolves well a very small compressibility, of the order of 1000 (eV)$^{-1}$, corresponding to 160aF, with about 1aF resolution. This is about 7 orders of magnitude smaller than the compressibility of a piece of metal of (1μm)$^3$. Remarkably, our sensitivity corresponds to a charge of about $2.5 \times 10^{-4}e$, which is about an order of magnitude lower than the charge sensitivity of an RF-SET setup[21] and 3 orders of magnitude lower than low frequency techniques[10,11].

The physics becomes strikingly different in the Kondo regime. For that purpose, we tune the gate of the device to $V_g \sim 2.5$V, where $\Gamma \sim 1$meV, and $E_C \sim 2.25$ meV. As shown in figure 3a, the conductance colour-scale plot displays softer Coulomb diamonds with horizontal Kondo ridges close to zero bias. The observation of several adjacent Kondo ridges is consistent with previous observations in carbon nanotubes[16]. It arises from the existence of additional degeneracies besides the spin in the spectrum of the nanotube. From the width of the zero bias peaks, we can estimate a Kondo temperature of about 5K. The main result of this paper is presented in figure 3b. Whereas there is a finite zero bias peak in the conductance (and therefore in the density of states of the dot), the simultaneously measured phase contrast shows that this density of states does not contribute to the compressibility (see Extended Data Figure 3 for the complete compressibility map). Importantly, the high energy charge peaks at about +/-2mV remain both in the phase and in the conductance, although they do not fully coincide. These experimental results are robust since we observe them for all the Kondo ridges studied (we present 15 examples in Extended Data Figure 7). Our experimental findings are well reproduced by Numerical Renormalisation Group (NRG) calculations of the $\chi$ and the dot density of states $\nu(E)$. The latter can be directly mapped onto the conductance $G(eV_{sd})$, plotted in figure 3d, by making the identification $E=eV_{sd}$, thanks to the small height of the Kondo peak (~0.12 $2e^2$/h) which ensures that the dot is much more tunnel coupled to one of the two reservoirs. The low bias data directly show that a finite (DC) current flows through the device, although the charge in the QD is frozen[4,5,6]. One can explain this feature within the Kondo model, as illustrated in figure 3c. When a QD degenerate level is singly occupied by a frozen charge, an antiferromagnetic coupling appears between the single electron and the conduction electrons at the Fermi energy. The emergent many body state does not contribute to the compressibility, because it arises from virtual tunnelling processes. Therefore, our measurements strongly suggest that the Kondo resonance in the conductance, also called Abrikosov-Suhl-Nagaoka resonance, is associated to the fluctuations of the spin degree of freedom whereas the charge fluctuations in the dot are frozen. Interestingly, the NRG data is also able to reproduce the shift between the conductance and compressibility charge peaks around +/- 2mV. We speculate that this might be a correlation effect related to an interaction-driven renormalization of the system parameters.

The temperature dependence of the cavity and transport signals further confirms that the conductance and the compressibility obey different physical principles governed by different energy scales. When the temperature increases, the many-body Kondo resonance decreases logarithmically on a temperature scale set by $T_K$, as shown in figure 4a. The residual compressibility $\chi_V$ in the valley evolves on a different temperature scale than the many-body Kondo resonance, as it is simply due to single electron tunnelling and set by $\Gamma$ ($\chi_V \sim -0.13/(\pi\Gamma)$ at $T=0$ from the NRG data). A linear fit to the data plot in log-linear scale in the high temperature range gives a logarithmic law of about $-0.18\,Log(T/T_K)$ for the conductance and of about $-0.73\,Log(T/\Gamma)$ for the compressibility. In figure 4b, we show the corresponding plots obtained by NRG. We find that they are in good agreement with the experimental data. In particular, the NRG data in figure 4b indicate that the temperature dependence for the conductance is governed by $T_K$ whereas for the compressibility it is governed by $\Gamma$. It is important to note however that extracting accurately the value of $\Gamma$ from our experimental data is not straightforward here because the apparent spectral (dI/dV) width of the charge resonance has been observed to depend on interaction (see Methods). This is also seen in the NRG data in figure 3d. This can explain why the temperature scale for the down-turns for the conductance and the compressibility are less separated in our experimental data than in the NRG data. Nevertheless, both the distinct slopes and the separate down-turns show that the conductance and the compressibility are affected by temperature with different mechanisms. This directly stems from the decoupling of the charge and spin degrees of freedom in a Kondo cloud.

In conclusion, we have directly observed the freezing of charge dynamics which is a crucial feature of a Kondo resonance. Our dual conductance/compressibility measurements illustrate the fundamental difference between a Kondo resonance and a simple resonant level where many body effects are absent.

Our setup can be generalized to many types of mesoscopic circuits[22,23,24] and could be transposed in the optical domain to probe the compressibility of other types of conductors. It could be used to study in a controlled manner some important fermion-boson problems. Electron-phonon interactions in solids could be simulated by using the analogy between phonons and the photons in our cavity. Furthermore, the cavity photons are slow here with respect to the electrons of the dot ($hf_{cav} \ll E_C, \Gamma, k_B T_K$), a situation that has allowed us to probe non-invasively the low frequency charge dynamics of the QD, relevant to understand the DC properties of our system. We expect to access dynamical aspects of tunnelling[14,25] and Kondo physics if one of these inequalities is not fulfilled. Among the perspectives offered by our findings, one could also imagine to inject suddenly a coherent field in the cavity to perform a quantum quench of the system which could give interesting insights into the dynamics of the Kondo cloud.

**Supplementary Information** Supplementary Information accompanies the paper on www.nature.com/nature.

**Acknowledgements.** We are indebted with L.I. Glazman, H. Baranger and A. Clerk for discussions. We gratefully acknowledge L.C. Contamin, T. Cubaynes, Z. Leghtas and F. Mallet for a critical reading of the manuscript and J. Palomo, M. Rosticher and A. Denis for technical support. The devices have been made within the consortium Salle Blanche Paris Centre. We acknowledge support from Jeunes Equipes de l'Institut de Physique du Collège de France (JEIP). This work is supported by ERC Starting Grant CIRQYS and by the NRF of Korea (Grant Nos. 2009-0069554 and 2011-0030046 for ML and 2015-003689 for MSC).

**Authors contributions.** MMD setup the experiment, made the devices and carried out the measurements with the help of TK. MMD performed the analysis of the data with inputs from TK. JJV, MCD LEB and MRD contributed through early experiments and to develop the nanofabrication process. MCD developed the data acquisition software. ML and MSC carried out the NRG calculations. MMD, TK and AC carried out the semi-classical theory of dot-cavity coupling. TK, AC and MMD co-wrote the manuscript with inputs from all the authors.

**Author Information.** Reprints and permissions information is available at www.nature.com/reprints. The authors declare no competing financial interests. Correspondence and requests for materials should be addressed to T.K. : kontos@lpa.ens.fr

**Figure 1 | Hybrid quantum dot-cavity QED setup**

**a.** The compressibility of an electron gas is associated to the ratio $\partial N/\partial \mu$, with $N$ the mean number of electrons and $\mu$ the chemical potential. **b.** The finite compressibility $\chi$ of the electron gas shifts the resonance frequency of a microwave cavity by $g^2\chi$ (ON state) from its bare resonance frequency $f_{cav,bare}$ (OFF state). The phase of the transmitted microwave signal at $f_{cav,bare}$ is thus shifted by $\Delta\phi$. The constant $g$ is the electron-photon coupling and $\kappa$ is the cavity linewidth. **c.** A carbon nanotube based quantum dot circuit is capacitively coupled to a coplanar waveguide microwave cavity. The chemical potential of the dot is controlled by the gate voltage $V_g$. The source-drain bias $V_{sd}$ is applied between the two electrodes (in blue) which delimit the quantum dot.

**Figure 2 | Nature of the electron-photon coupling**

**a.** and **b.** Conductance G and phase maps in the $V_g$-$V_{sd}$ plane for low gate voltages. The opposite phase is represented in order to map directly the microwave signal onto the compressibility. **c.** Gate sweep for $V_{sd}$~0 for the conductance (top panel) and for the phase (bottom panel). The points are experimental data and the solid lines correspond to lorentzian fits. **d.** Bias sweep at $V_g$=1.33V. **e.** Coupling mechanism: the cavity photons modulate adiabatically the chemical potential of the quantum dot. The dot has tunnelling rates $\Gamma_S$ and $\Gamma_D$ to the source S and the drain D, respectively. A finite dot density of states at the Fermi level $E_F$ turns on electronic transfers between the quantum dot and the leads. This dipole induces a shift in the resonant frequency of the cavity (top panel), which leads to the phase shift seen in b, c and d.

**Figure 3 | Transparent Kondo resonance.**

**a.** Conductance map in the Kondo regime. **b.** Simultaneous bias dependence of the conductance and the phase in the middle of a Kondo ridge along the black dashed line. The blue (orange) arrows mark the charge peaks in the conductance (compressibility). The black arrow marks the Kondo resonance. **c.** A quantum dot level away from the Fermi energy leads to a Kondo resonance through a sum of virtual processes (dashed line). **d.** Numerical renormalization group (NRG) data corresponding to the situation of panel b. The excitation energy $E$ is scaled by the charging energy and the compressibility by $\pi\Gamma$. Both the absence of Kondo peak in the compressibility as well as the shift of the high energy charge peaks between compressibility and conductance are reproduced. The corresponding parameters are $\Gamma=0.4E_C$ (as extracted from the data) and $T = 10T_K$.

**Figure 4 | Temperature dependence of conductance and compressibility.**

**a.** Temperature dependence of the conductance and the phase on the Kondo ridge at ($Vg \sim 2.567\,V,\ Vsd \sim -0.15mV$) (see Methods and Extended Data Figure 6). In order to compensate thermal drifts, the phase shift signal ΔPhase on the Kondo ridge is measured with respect to the right adjacent Coulomb peak at a given temperature. The error bars are about the size of these experimental points. The dashed lines show linear fit in log scale that corresponds to a logarithmic law $-\alpha Log(T)$. **b.** NRG data as a function of temperature, for $\Gamma=0.4E_C$ (as extracted from the data).

**Extended Data Figure 1 | Microwave cavity characterization.** Left panel: Phase and amplitude of the microwave signal as a function of frequency

showing the cavity resonance used to measure the compressibility. Right panel: temperature dependence of the linewidth of the cavity.

**Extended Data Figure 2 | Coulomb blockade regime.** Phase and conductance on wide scale in Coulomb blockade regime. The observation of groups of four peaks both in the conductance and in the phase contrast arises from the spin/valley degeneracy of the nanotube spectrum.

**Extended Data Figure 3 | Phase in the Kondo regime.** Colorscale plot of phase in the Kondo regime corresponding to figure 3a in the main text. We observe tilted lines arising from single charge peaks but no Kondo ridge. The tilted doted black lines are guides to the eye. The vertical dashed lines correspond to the position of the cuts presented in the main text (first), and in the Methods section (third for Extended Data Figure 6 left panel and second and forth for Extended Data Figure 7). A spurious titled blue line is also observed. It likely arises from an impurity level coupled to the cavity field.

**Extended Data Figure 4 | Systematics for the Kondo regime. a** and **b.** Conductance and phase as a function of source-drain bias and gate voltage for different Kondo ridges than the set presented in the main text. **c**. Conductance and phase as a function of source-drain bias and gate voltage on a wide scale in the Kondo regime. The measurements have been performed for a different cool-down (from 2K to 250mK) of our $^3$He single shot cryostat and correspond to different physical parameters than for panels a and b.

**Extended Data Figure 5 | Temperature dependence for other Kondo ridges. a.** Conductance (top panel) and phase (bottom panel) as a function of temperature for second Kondo ridge of figure 3a in the main text. **b.** Conductance (top panel) and phase (bottom panel) as a function of temperature for forth Kondo ridge of figure 3a in the main text.

**Extended Data Figure 6 | Kondo peak for temperature dependence.** Left panel: Bias dependence of conductance and phase for the Kondo ridge used to determine the temperature dependence of figure 4a. Right panel: Corresponding gate dependence at base temperature (255mK) and at high temperature (2.05K). To get rid of thermal drift of the phase, we compute the difference of the phase between a Coulomb peak (green arrow) and a Coulomb valley (blue arrow), where the Kondo ridge is. The phase at 2.05K has been rescaled to take into account the decrease of the quality factor with the temperature (22 000-> 18 000).

**Extended Data Figure 7 | Dual conductance/compressibility measurements for other Kondo ridges.** Examples for 15 different Kondo ridges displaying the same observation as in the main text. These data correspond to cuts indicated by vertical dashed lines in Extended Data Figure 4. In particular, the Kondo peak apparent in the conductance (in blue) is always absent from the compressibility (in orange).

**Extended Data Figure 8 | Control experiment for calibration of electron-photon coupling.** Power dependence of Coulomb peaks for 4 different peaks (a,b,c and d). Each peak height is plotted on the right panels versus the microwave modulation amplitude which controls the number of photons inside the cavity. The open dots are data and the solid lines are fits using formula (13).

**METHODS**

**Fabrication of the devices and measurement techniques.** A 150nm thick Nb film is first evaporated on an RF Si substrate at rate of 1nm/s and a pressure of $10^{-9}$ mbar. The cavity is made subsequently using photolithography combined with reactive ion etching ($SF_6$ process). An array of bottom gates is then made with two e-beam lithography steps in a 100μm square opening of cavity ground plane near the central conductor. First, we etch 750nm x 25 μm trenches of 130nm depth with reactive ion etching ($CHF_3$ process). Second, we deposit inside the trenches 150nm narrower layers of Ni(100nm)/AlOx(6nm).

The Al oxide is obtained by 3 steps of static oxidation of 2nm-thick Al layers using an O$_2$ pressure of 1 mbar for 10 min. Carbon nanotubes are grown with Chemical Vapor Deposition technique (CVD) at about 900°C using a methane process on a separate quartz substrate and stamped above the bottom gates[15]. The nanotubes are then localized and those which correctly lie on a bottom gate are contacted with Pd(4nm)/Al(80nm). During this last e-beam lithography and evaporation step, gate electrodes are also patterned in order to couple capacitively the bottom gate to a DC gate voltage V$_g$ and to the AC potential of the central conductor of the cavity.

The DC measurements are carried out using standard lock-in detection techniques with a modulation frequency of 77 Hz and an amplitude of 30 µV. The base temperature of the experiment is 255 mK. The microwave measurements are carried out using room temperature microwave amplifiers with a total gain of 60 dB. We measure both quadratures of the transmitted microwave signal using an I-Q mixer and low frequency modulation at 2.7 kHz. The cavity resonance frequency is 6.67129 GHz and its quality factor is between 10 000 and 20 000 depending on the run of our single shot $^3$He cryostat (see Extended Data Figure 1). The input power for the cavity is -89 +/- 2 dBm resulting in an average photon number of about 30000. This power yields a microwave modulation of about 40 µV which ensures that we are in the linear regime (consistent with the power dependence of the DC conductance; see below).

**Link between the phase shift of the microwave signal and the compressibility.** The transmission of the cavity in the frequency domain is shifted by $\hbar g^2 \chi / 2\pi$, where $\chi$ is the compressibility and $g$ the electron-photon coupling constant (see below):

$$T(f) = \frac{-i\kappa/2}{2\pi(f - f_{cav}) + \frac{i\kappa}{2} - \hbar g^2 \chi}$$

where $\kappa$ is the linewidth of the cavity.

We measure the transmission at resonance $f = f_{cav}$. This yields:

$$T(f) = \frac{A_{out} e^{i\Delta\varphi}}{A_{in}} = \frac{-i\kappa/2}{\frac{i\kappa}{2} - \hbar g^2 \chi}$$

Hence, for small phase shifts,

$$\Delta\varphi \simeq 2g^2 \hbar \chi / \kappa$$

This expression holds for any electronic system as long as the linear and adiabatic regime are reached. The parameter $\kappa \sim 2\pi \times 0.3$ MHz, yielding a quality factor Q= $f_{cav}/\kappa$ =18 000, can be measured directly from the transmission spectrum of the cavity (Figure 1b). At low temperatures, in the simplest case, the compressibility reads: $\chi = \frac{\partial N}{\partial \varepsilon_d} = -\frac{4}{\pi} \frac{\Gamma}{\Gamma^2 + 4\varepsilon_d^2}$, where $\Gamma$ and $\varepsilon_d$ are respectively the line width of the Coulomb peak and the position of the dot energy level. The dot's parameter $\Gamma = \Gamma_S + \Gamma_D$, with $\Gamma_S = 0.7$ meV and $\Gamma_D = 4\mu$eV can be determined from the conductance measurements (see figure 2) which also allow us to extract the charging energy $E_C$=3.5 meV. As a consequence, the joint conductance / phase measurements presented in figure 2c allow us to directly determine the electron-photon coupling constant g on each Coulomb peak, g~$2\pi \times$ (65MHz+/- 15MHz). The negative sign observed for the phase contrast shows directly that the dot reduces the frequency of the resonator. Therefore, the effective admittance of the QD circuit is that of an effective capacitance in parallel with the capacitance of resonator. This stems from the fact that cavity photons are coupled to the gate (and therefore $\varepsilon_d$), but not to the source-drain contacts. This feature of our setup is crucial to ensure that we measure only the compressibility of the electron system in the quantum dot, which was not the case in a previous experiment in the Kondo regime[17] (See below for an extensive discussion)

The large coupling strength found is consistent with our circuit design, shown in figure 1c, where a bottom gate (in green) very close to the single wall carbon nanotube combines the AC voltage of the central conductor of the cavity and the DC gate voltage (see below). The electron-photon coupling constant simply reads: $g = e\, \alpha\, V_{rms}$ (e>0), where $V_{rms}$ is the root mean square voltage associated to a single photon and $\alpha$ the ratio of the induced RF oscillations of the dot chemical potential to the potential of the central conductor. From the conductance map, we can infer the DC lever arm to about 0.3. From our gate layout (see figure 1c), we can estimate that the AC capacitance is about 3 times smaller than the DC capacitance. This leads to $\alpha \sim 0.1$, so that $g \sim 2\pi \times 50 MHz$, using Vrms ~ 2μV. This order of magnitude is in good agreement with our experimentally determined coupling strength of about $2\pi$ x 50-$2\pi$ x 100 MHz.

**Conductance and microwave phase in the Coulomb blockade regime on a wide gate voltage range.** The cavity resonance which is used to perform our compressibility measurements is presented in Extended Data Figure 1 left panel. We show in Extended Data Figure 2 the phase contrast and the conductance as a function of the gate voltage at zero bias in the Coulomb blockade regime on a wide gate voltage range. The conductance (in blue lines) displays regularly spaced Coulomb peaks with the expected fourfold periodic shell filling of low disordered single wall carbon nanotubes. The corresponding phase of the transmitted microwave signal exhibits a pattern which is very well correlated to the conductance. The phase contrast ranges from 1° to 5° probably due to modifications of the dot electronic wavefunctions due to weak disorder. Nevertheless, the extracted value for the coupling strength remains of about $2\pi$ x 50-$2\pi$ x 100 MHz.

**Phase colorscale plot in the Kondo regime.** We show in Extended Data Figure 3 the phase colorscale plot measured simultaneously with the conductance colorscale plot shown in the main text in figure 3a. Similarly to the Coulomb blockade regime (figure 2b), the phase contrast is bigger for the left tilted edges of the Coulomb diamonds (the tilted black lines are guides to the eye). The latter are blurred as expected since the Kondo regime corresponds to a gate region with larger $\Gamma$'s. There are some phase resonances which are not correlated to the Coulomb diamond edges like in the area around (Vsd=-2mV, Vg=2.54V). We attribute this effect to spurious impurity levels which are coupled to the cavity. Importantly, the Kondo ridges which are clearly visible in the colormap of the conductance are completely absent from the phase map.

**Phase and conductance for additionnal Kondo ridges.** We show in this section the robustness of our findings. We present the dual conductance and compressibility cuts as a function of Vsd for 15 Kondo ridges in Extended Data Figure 7. These cuts correspond to the dashed lines represented on the colorscale plots of Extended Data Figure 4 (for 13 of them). The position of the cuts for the $2^{nd}$ and the $4^{th}$ ridge is shown in Extended Data Figure 3 which corresponds to the phase contrast for figure 3a of the main text are also presented in this panel of 15 cuts. Essentially all what is described for the Kondo rigde of the main text is observed. The Kondo peak is present in the conductance (in blue) but not in the compressibility (in orange). This further confirms the robustness of our findings.

We also present in Extended Data Figure 5 the temperature dependence for the joint conductance and compressibility for the 2nd and 4th ridge of the figure 3a of the main text. The slope for the 4th conductance ridge is found to be around -0.15 (panel a top of Extended Data Figure 5), close to the value for the 3rd ridge third ridge presented in the main text. The slope for the 2nd conductance ridge (panel b top) is more difficult to estimate as can be seen from the spread of the data points probably arising from small gate drifts as we increase the temperature. This is even more difficult for the data of the 1st ridge (corresponding to the cut presented in the main text) and this is why we do not present it. Both the compressibility in panel a and b is shown to start to decrease after the down-turn in the conductance as highlighted by the blue and orange regions. This is clearer in panel a than in panel b probably due to the larger spread in the compressibility measurements in panel b. Note that both slopes for compressibility are different than what is presented in the main text. This is expected since the slope for the compressibility is non universal and depends on $\Gamma$.

**Phase and conductance for Kondo ridge used in figure 4.** We show in Extended Data Figure 6 the dual conductance/compressibility measurement for the center of the third Kondo ridge in figure 3a. The same features as in figure 3 are observed. While a peak is visible in the conductance, signalling the Kondo resonance, it is absent from the compressibility. The charge peaks are still visible in both measurements, around +/- 2mV. As for the example of the main text, they are not fully correlated.

In order to obtain the temperature dependence of figure 4a, we use gate scans for Vsd=0mV at different temperatures as shown in Extended Data Figure 6 right panel. As described in the main text, the right adjacent Coulomb peak is used as a reference for the phase. In addition, in order to get a meaningful temperature dependance of the phase, we rescale the data by the relative variations of the quality factor of the cavity measured at each temperature (see Extended Data Figure 1 right panel for the temperature dependence of the linewidth of the cavity).

**Theory of cavity-quantum dot coupling in the adiabatic regime.** We present in this section the general theory describing the cavity response in the presence of a quantum dot with various light-matter coupling schemes. The most general hamiltonian describing the hybrid quantum dot cavity system is[26]:

$$\hat{H} = \hbar\,\omega_{cav}\,\hat{a}^\dagger\,\hat{a} + (\hbar g\,\hat{n} + \hbar g_S\,\hat{n}_S + \hbar g_D \hat{n}_D)\,(\hat{a} + \hat{a}^\dagger) + \hat{H}_{dot} + \hat{H}_{tunnel\ dot/lead} + \hat{H}_{Baths}$$

where $\omega_{cav}$ is the pulsation of the cavity and $\hat{a}^\dagger(\hat{a})$ the creation (annihilation) operators for the cavity photons and $\hat{n}$ the number of electrons in the dot. The coupling term includes the charge on the dot characterized by the operator $\hat{n}$ as well as the number of electrons in the source (S) and drain (D) reservoirs characterized by operators $\hat{n}_{S(D)}$. As we will see below, the existence of these different couplings have specific consequences on the cavity signals. It is therefore possible to infer which term dominates in order to demonstrate that the compressibility of the dot is directly measured.

It is now useful to define $\hat{n}_{+/-} = \hat{n}_S \pm \hat{n}_D$. Charge conservation imposes the conservation of $\hat{n}_+ + \hat{n}$. We consider an excitation of the cavity with the form : $\langle \hat{b}_{in} \rangle = b_{in}^0(t)e^{-i2\pi ft}$, with $f \approx f_{cav}$. There are a few thousands of photons in the cavity so we can use the semi-classical approximation for the photonic field. The equation of motion of the amplitude $\langle \hat{a} \rangle = \bar{a}(t)e^{-i2\pi ft}$ of the field in the semiclassical limit reads :

$$\dot{\bar{a}} = [i2\pi(f - f_{cav}) - \kappa/2]\,\bar{a} - ig_+ N - ig_- N_- - \sqrt{\kappa_{in}}\,b_{in} \qquad (1)$$

with $N = \langle \hat{n} \rangle$, and $N_- = \langle \hat{n}_- \rangle$, $g_+ = g - \frac{g_S + g_D}{2}$ and $g_- = \frac{g_S - g_D}{2}$. In the stationary regime, the cavity field reads[2] :

$$\bar{a} = \frac{-i\sqrt{\kappa_{in}}\,b_{in}}{2\pi(f - f_{cav}) + i\frac{\kappa}{2} - \hbar \sum_{ij \in \{+,-\}} g_i g_j \chi_{ij}(f)} \qquad (2)$$

The susceptibilities $\chi_{ij}(f)$ are the Fourier transforms of $\chi_{ij}(t) = -i\theta(t)\langle[\hat{n}_i(t), \hat{n}_j(0)]\rangle$: $\chi_{ij}(f) = \int_{-\infty}^{+\infty} dt\,\chi_{ij}(t)e^{i2\pi ft}$ \qquad (3)

In general, all these susceptibilities depend on the frequency. However, when the frequency of the cavity is much smaller than all the characteristic frequencies involved in the dynamics of the dot (essentially $\Gamma_{S(D)}$ in the single dot case), one may use the adiabatic limit of $\chi_{ij}(f)$ i.e. $\chi_{ij}(f \to 0)$. Since, $\hat{n}$ couples to the energy level $\epsilon_d$ and $\hat{n}_-$ couples to the source-drain bias and is $\frac{2I}{-i2\pi f e}$, where $I$ is the average current, one may write :

$$\sum_{ij \in \{+,-\}} g_i g_j \chi_{ij}(f)$$

$$= g_+^2 \frac{\partial N}{\partial \epsilon_d} + g_-^2 \frac{4}{e^2} \frac{1}{i2\pi f} \frac{\partial I}{\partial V_{ac}} - 2 g_- g_+ \frac{\partial N}{\partial e V_{ac}}$$

$$+ 2 g_+ g_- \frac{1}{e} \frac{1}{i 2\pi f} \frac{\partial I}{\partial \epsilon_d} \qquad (4)$$

with $N = \langle \hat{n} \rangle$. For $V_{sd} = 0$, the current I is zero independently of $\epsilon_d$, which implies that the last term in the above equation is zero. The quantities $\frac{\partial I}{\partial V_{ac}}$ and $\frac{\partial N}{\partial V_{ac}}$ are the derivatives of the current and the charge of the dot when the bias is applied symmetrically since the perturbation is proportional to $\hat{n}_-$. The second term is purely imaginary since $\frac{\partial I}{\partial V_{ac}}$ is real. However, one can see that $\frac{\partial N}{\partial V_{ac}}$ can also contribute to the real part of the response and therefore to the phase shift with prefactor $g_- g_+$. It is therefore important to assess that the dominant response comes from the $g_+^2 \frac{\partial N}{\partial \epsilon_d}$ term which corresponds to $g_+^2 \chi$, where $\chi$ is the compressibility defined in the main text. For that purpose, we use the second (dissipative) term and the fact that its derivation is valid for any circuit in the adiabatic regime. The Kondo peak is absent from the amplitude data (not shown) to the experimental uncertainty of 0.001: this implies that $g_- < 2.0 MHz$ for example for the peak shown in Extended Data Figure 5. This is more than an order of magnitude smaller than $g_+$.

It is instructive to evaluate the general form of the dot response in the adiabatic and non-interacting regime. We may define:

$$\chi_0(\epsilon_d) = -\int \frac{d\epsilon}{2\pi} \mathcal{A}(\epsilon - \epsilon_d) \frac{1}{4k_B T \cosh^2\left(\frac{\epsilon}{2k_B T}\right)} \tag{5}$$

where $\mathcal{A}(\epsilon - \epsilon_d)$ is the density of states of the dot. Using for example the Keldsyh formalism, we get :

$$\frac{\partial N}{\partial \epsilon_d} = \Gamma_S \chi_0(\epsilon_d - eV_S) + \Gamma_D \chi_0(\epsilon_d - eV_D) \tag{6}$$

$$\frac{\partial N}{\partial \epsilon_a} = -\Gamma_S \chi_0(\epsilon_d - eV_S) + \Gamma_D \chi_0(\epsilon_d - eV_D) \tag{7}$$

$$\frac{\partial I}{\partial \epsilon_d} = \Gamma_S \Gamma_D [\chi_0(\epsilon_d - eV_S) - \chi_0(\epsilon_d - eV_D)] \tag{8}$$

$$\frac{\partial I}{\partial \epsilon_a} = \Gamma_S \Gamma_D [-\chi_0(\epsilon_d - eV_S) - \chi_0(\epsilon_d - eV_D)] \tag{9}$$

Where $\epsilon_a$ correspond to the anti-symmetric modulation $V_S = -V_D = \frac{\epsilon_a}{-e} = \frac{V_{ac}}{2}$.

The above expressions show the equivalence between a compressibility measurement and a conductance measurement in the non-interacting regime.

**Alternative scenario for the 'transparent' Kondo resonance.** In the Kondo regime, we have seen that the conductance peak at zero bias is absent from the cavity signal.

Could this conductance peak be attributed to another level, pinned at zero bias, with a chemical potential $\epsilon'_d$ which is not coupled to the cavity?

From the bias dependence, this level has a width $\Gamma' = k_B T_k$. Its conductance appears and disappears at some gate voltage $V_g$. The conductance depends only on the chemical potential and $\Gamma'$ of the level. As the chemical potential is not coupled to the cavity, the gate has also no influence on it, so $\frac{\partial \epsilon'_d}{\partial V_g} \sim 0$. Therefore, the level should have at least one of its $\Gamma'$, for example $\Gamma'_D$, that depends on the gate : $\frac{\partial \Gamma'_D}{\partial V_g} \neq 0$.

The conductance appears as a lorentzian with respect to the bias voltage $V_{SD}$, so the charge number $N'$ is :

$$N' = \frac{1}{2} - \frac{2}{\pi} \arctan \frac{2eV_{SD}}{\Gamma'_S + \Gamma'_D} \qquad (10)$$

Its derivative with respect to the gate is then :

$$\frac{\partial N'}{\partial V_g} = \frac{4}{\pi} \frac{\Gamma'_S + \Gamma'_D}{4(eV_{SD})^2 + (\Gamma'_S + \Gamma'_D)^2} \frac{2eV_{SD}}{\Gamma'_S + \Gamma'_D} \frac{\partial \Gamma'_D}{\partial V_g} \qquad (11)$$

For a coupling to the dot chemical potential, we have $\frac{\partial N}{\partial V_g} = \frac{\partial \epsilon_d}{\partial V_g} \frac{\partial N}{\partial \epsilon_d} = \alpha \chi$, where $\chi$ is the compressibility and $\alpha$ the lever arm. By analogy, $\frac{\partial \Gamma'_D}{\partial V_g}$ is the lever arm $\alpha'$ that enters in the coupling parameter $g'$, describing the response of the level to a modulation of $\Gamma'_D$. From the conductance data, one can estimate that this lever arm is 0.75 smaller than the lever arm $\alpha$ of the level that give the Coulomb peaks. Therefore the phase should be shifted at the end of each Kondo ridge by the same amount as for the Coulomb peaks. Our setup can therefore exclude such a situation without the need for extra knowledge on the system.

**Figure of merit of our compressibility measurements.** We describe in greater details than in the main text the figure of merit of our compressibility measurement setup. Two main features are important for defining the figure of merit: first, the effective capacitance resolution $\delta C$ which can be achieved and second, the maximum excitation voltage which is used for that measurement $\delta V$. The latter is crucial for keeping the linearity of our detection scheme. These two parameters enter into the charge resolution of the setup: $\mathcal{N} = \delta C \times \delta V/e$. In our case, since we estimate $\delta C$ ~1aF from our phase noise of about 0.01 degree and we estimate $\delta V$~40 µV from the average number of photons in our cavity, this leads to $\mathcal{N} \approx 2.5 \times 10^{-4} e$. As a comparison, the minimum $\delta C$ in ref 8 of the main text is 1aF but with a $\delta V$ of about 20 mV.

**Photon number dependence of the differential conductance in the Coulomb blockade regime.** In this section, we show that one can estimate the electron-photon coupling strength with a complementary method than that used in the main text, from the microwave power dependence of the conductance. For a coupling to the gate, and in the adiabatic case $f_{cav} \ll \Gamma$, the conductance is modulated by the cavity photons as :

$$G(t) = G\big(\epsilon_d + 2\hbar g\sqrt{\bar{n}}\cos(2\pi f_{cav}t)\big) \quad (12)$$

where $\bar{n}$ is the average photon number. The conductance is a lorentzian with a width $\Gamma$, hence, at $\epsilon_d = 0$, a DC measurement gives

$$G = \int_0^{2\pi} \frac{d\theta}{2\pi} G(2\hbar g\sqrt{\bar{n}}\cos(\theta)) = \frac{1}{\sqrt{1 + 16 * \bar{n} * (\hbar g/\Gamma)^2}} \quad (13)$$

The mean number of photon depends on the cavity input power and its transmission:

$$\bar{n} = \frac{10^{\frac{P_{IF}+S_{att}}{10}+\frac{S_{cav}}{20}-3}}{\pi\kappa h f_{cav}} \quad (14)$$

$P_{IF}$ is the power in dBm corresponding to the root mean square amplitude $V_{IF}$ of the low-frequency microwave modulation. $S_{att}$ is the attenuation of RF lines to the cavity, calibrated to -82 +/- 2dB and $S_{cav}$ the transmission of the cavity, -16 dB. This allows to calibrate the coupling constant g ~2π x 60 MHz, which is in good agreement with the one deduced from the height of the Coulomb peaks: $g^2 \simeq \frac{\pi}{4} * \Delta\varphi * \Gamma * \kappa/\hbar$, as shown in the right panels of Extended Data Figure 8.

**Numerical Renormalization Group (NRG) calculations.** We calculate the compressibility using the numerical renormalization group (NRG) Method[27-29]. We have adopted the Anderson impurity model:

$$H = \sum_{k\sigma} \epsilon_k c^\dagger_{k\sigma} c_{k\sigma} + \sum_\sigma \epsilon_d d^\dagger_\sigma d_\sigma + U n_\downarrow n_\uparrow + \sum_{k\sigma} t_k c^\dagger_{k\sigma} d_\sigma + h.c.$$

where $c^\dagger_{k\sigma}$ and $c_{k\sigma}$ are the annihilation and creation operators, respectively, of the conduction electrons with momentum k and spin σ, whose energy is $\epsilon_k$, and $d^\dagger_\sigma$ and $d_\sigma$ are the same operators for the electrons on the quantum dot, whose energy is $\epsilon_d$. U describes the Coulomb interaction on the dot and $n_\sigma = d^\dagger_\sigma d_\sigma$. Note that in the orthodox charging model U~2$E_C$ ($E_C = e^2/2C$ with C being the capacitance of the dot). In experiments, the tunneling amplitude $t_k$ is assessed through the level hybridization (or tunneling-rate) parameter: Γ

$(E) = \pi \sum_k |t_k|^2 \delta(E - \epsilon_k)$. Assuming sufficiently wide conduction band, the energy-dependence in $\Gamma(E)$ is ignored.

As the NRG method works only at equilibrium, we adopt the approximation $\chi(f \approx 0, V = V_{sd}) \approx \chi_{NRG}\left(E = \frac{eV_{sd}}{\hbar}, V = 0\right)$, which is reasonably good in the linear-response regime and static limit $f_{cav} \ll T_K$. Direct application of the dynamical NRG method[27,28] gives the imaginary part, $Im(\chi)$ of the zero-bias compressibility $\chi(E) \equiv \chi(E, V = 0)$, and the Kramers-Kronig relation yields the real part :

$$Re[\chi(E)] = -\frac{1}{\pi} Pr \int_{-\infty}^{+\infty} dE' \frac{Im[\chi(E')]}{E - E'} \qquad (15)$$

where Pr denotes the Cauchy principal value.

The NRG method divides the entire energy range into discrete sectors of the logarithmic scale, and integrates the high-energy sectors iteratively until the required low-energy sector is reached. In this iterative procedure, it is important to keep the same level of accuracy for the higher-energy sectors (earlier stage of the iteration) because we are interested in the high- energy regime ($E \sim \epsilon_d, U$) as well as the low-energy range ($|E| < k_B T_K$). To achieve this goal, we adopt the density-matrix NRG method[30,31], where the dynamical excitation spectral density is obtained from the reduced density matrix of each energy sector. In order to enhance the speed and efficiency in the sampling of the spectral peaks in the logarithmic energy scale, we have also used the so-called z-trick[32]. Typically we take the z-average over 32 different z values. In this NRG study, we have found two interesting high-energy properties that have been largely overlooked in previous studies (which mostly have focused on low-energy properties): (i) The charging peak at $E \approx \epsilon_d$ of the compressibility is shifted from that of the conductance by an amount comparable to $\Gamma$. This shift is clearly observable in the experimental result. (ii) The width of the charging peak (at $E \approx \epsilon_d$) of the conductance for U>> $\Gamma$ is almost twice wider than that (~$\Gamma$) for the non-interacting case (U = 0)[32]. This is also consistent with the value of $\Gamma$ when estimated from the experimentally measured dI/dV data.

**Data availability.** The authors declare that the main data supporting the findings of this study are available within the article (main text, methods and extended data). Extra data are available from the corresponding author upon request.

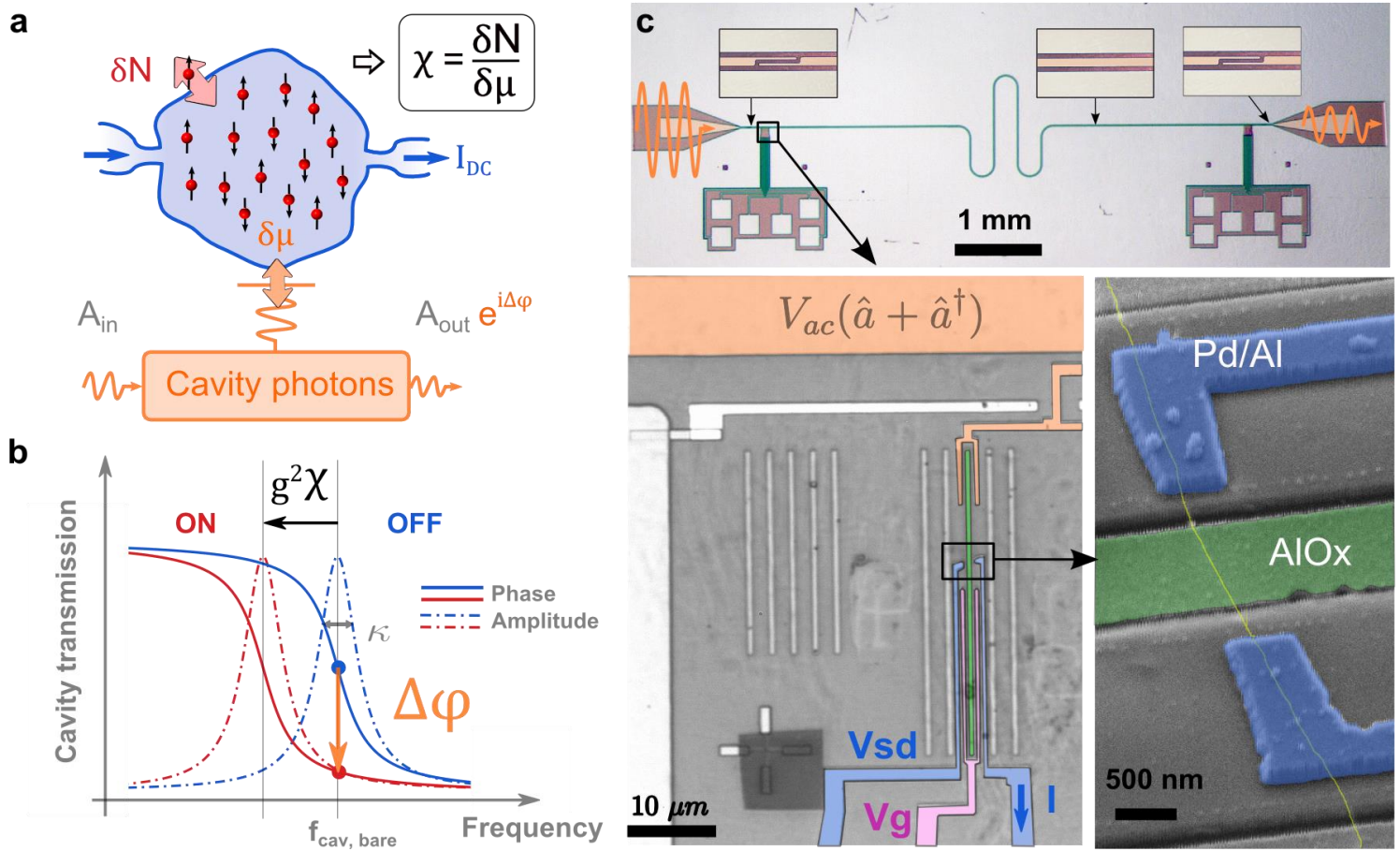

Fig. 1 Desjardins et al.

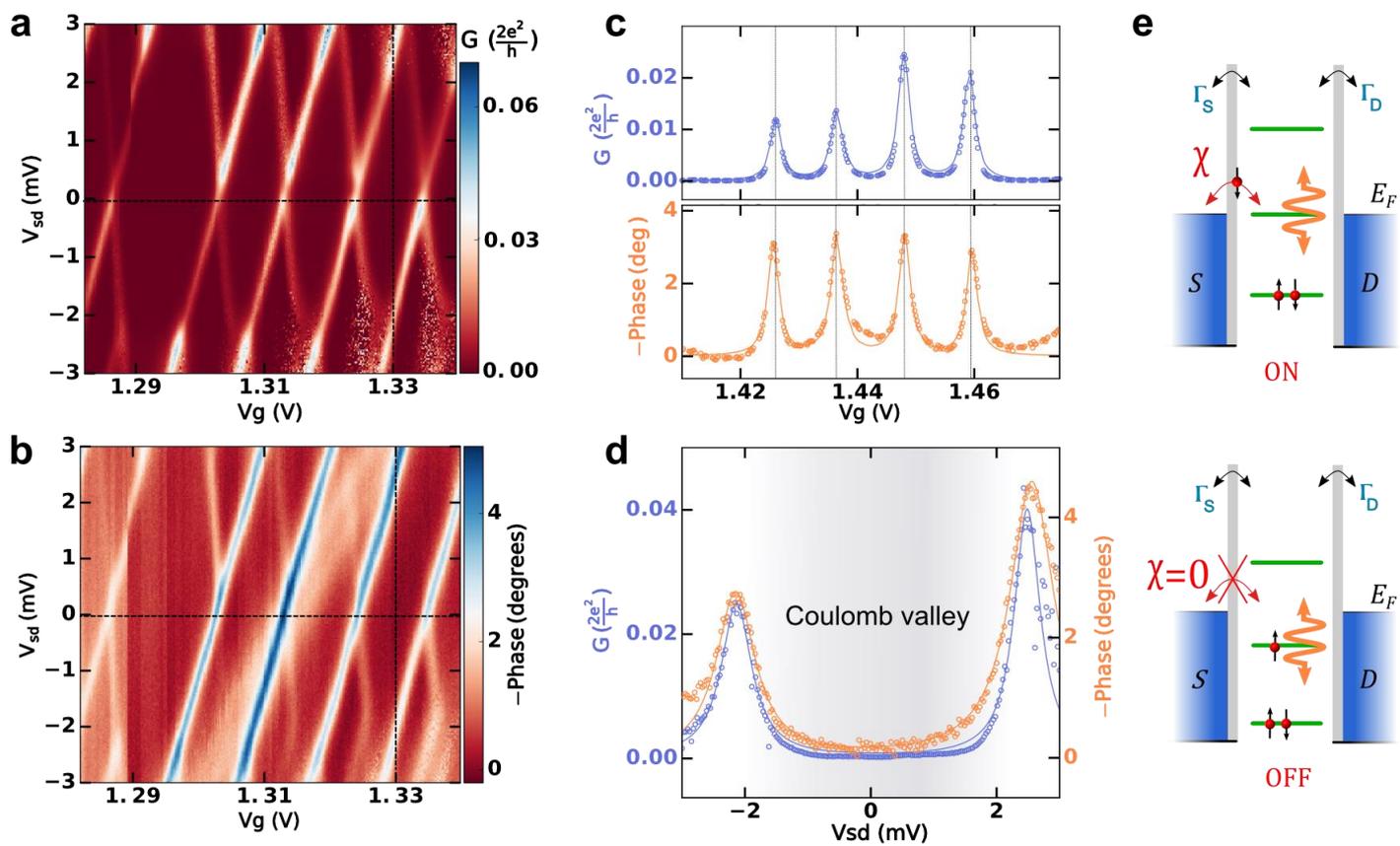

Fig. 2 Desjardins et al.

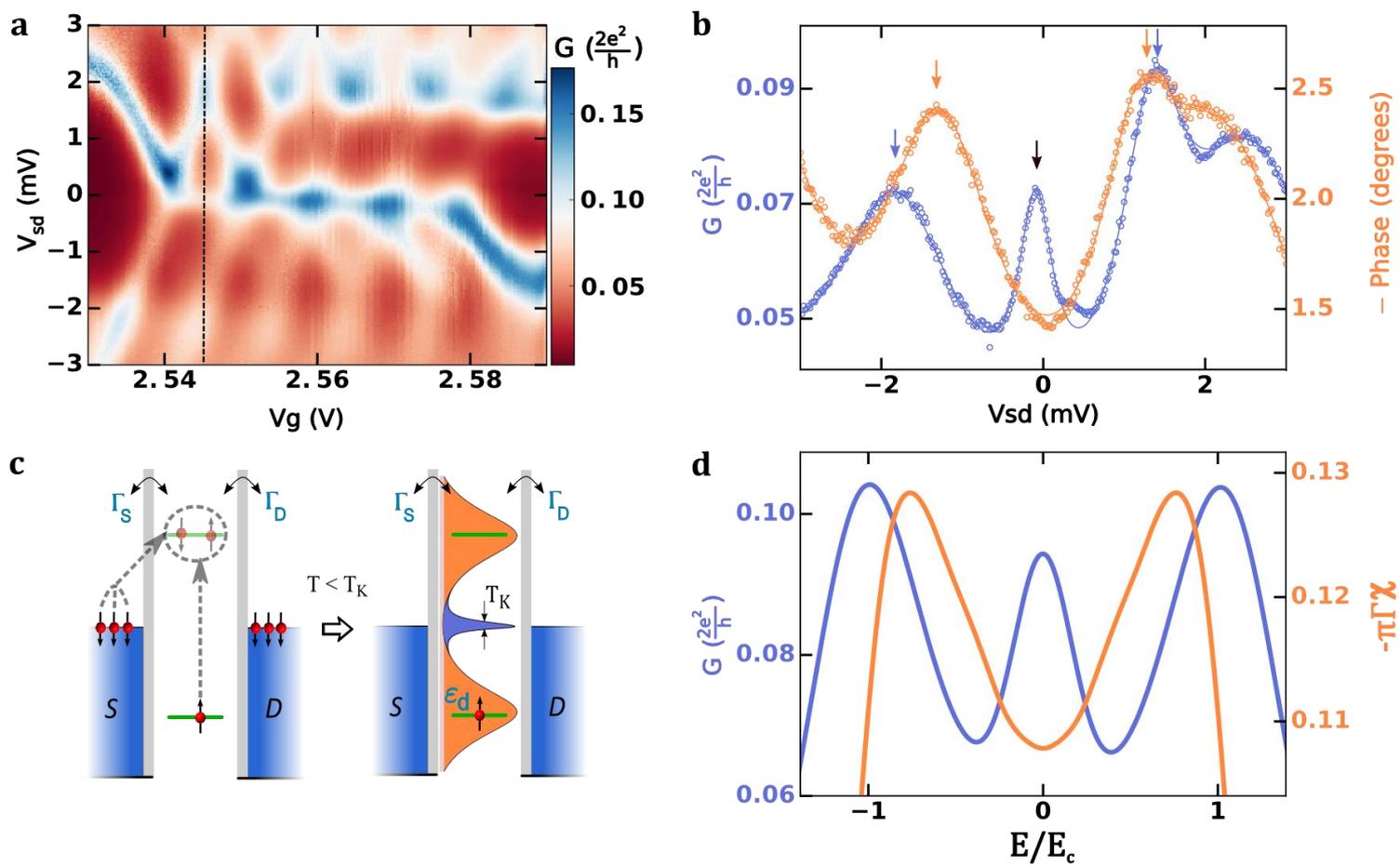

Fig. 3 Desjardins et al.

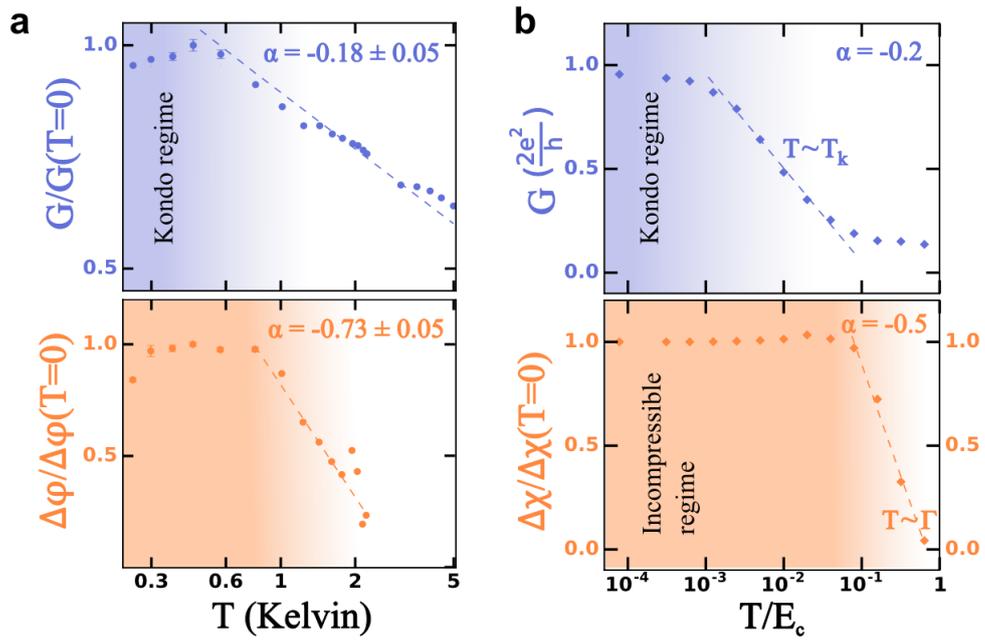

Fig. 4 Desjardins et al.

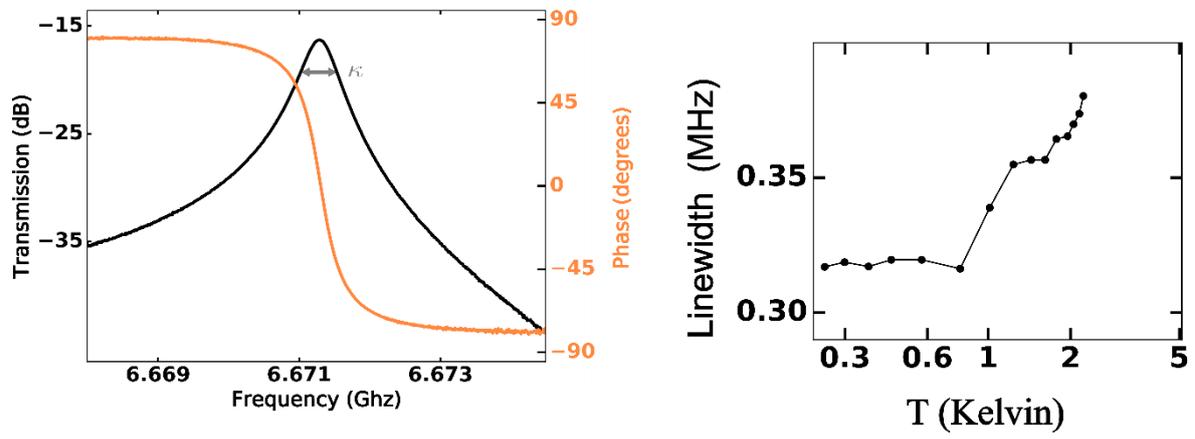

Extended Data Fig. 1
Desjardins et al.

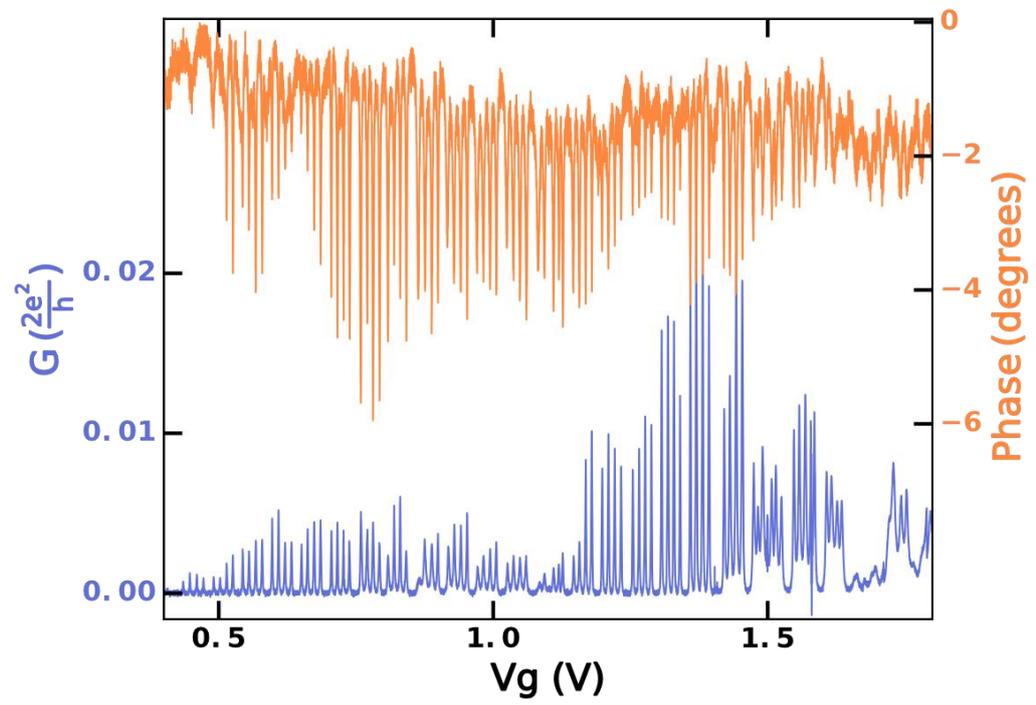

Extended Data Fig. 2
Desjardins et al.

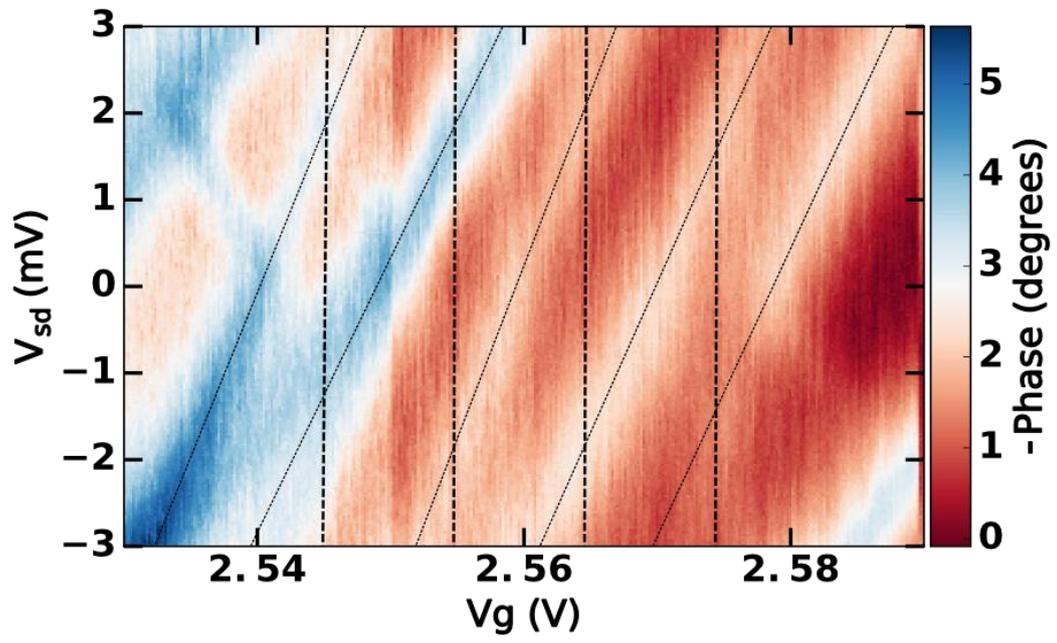

Extended Data Fig. 3
Desjardins et al.

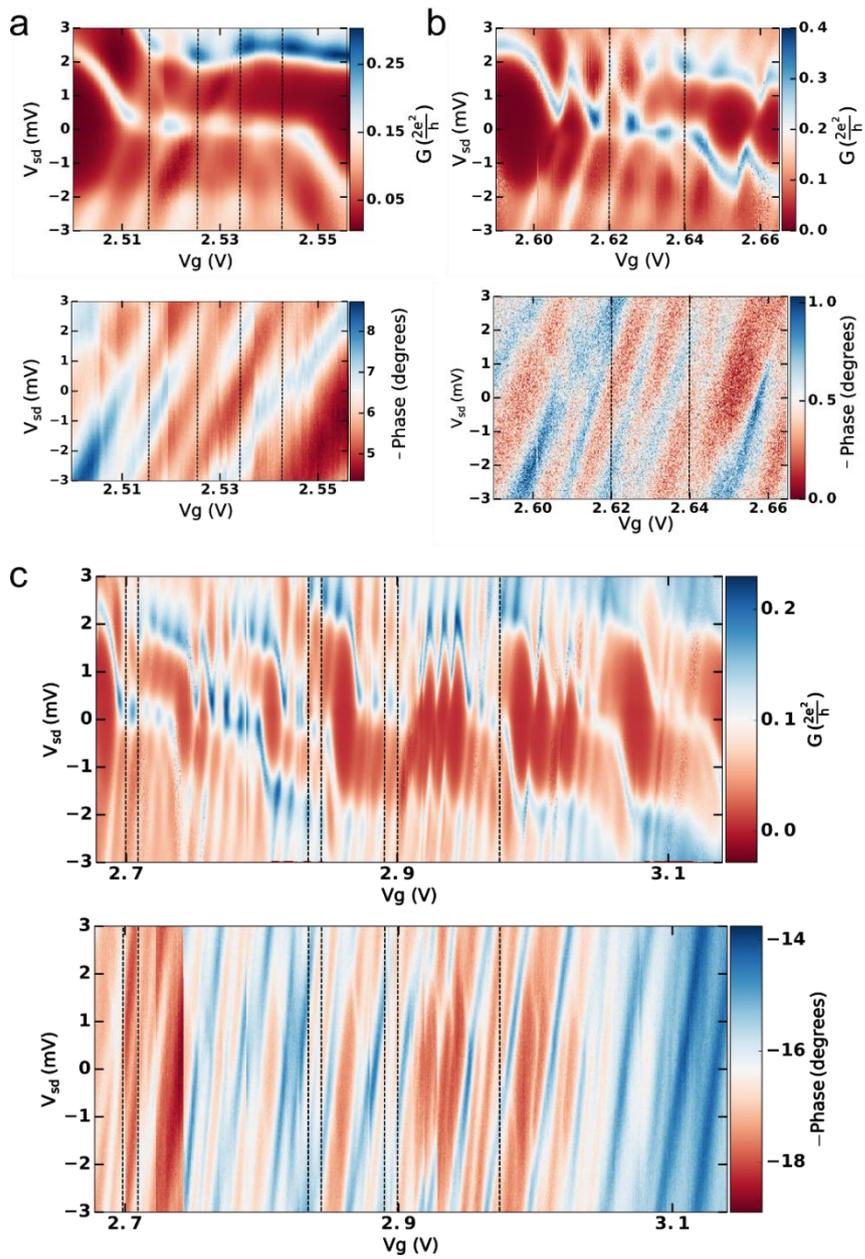

Extended Data Fig. 4
Desjardins et al.

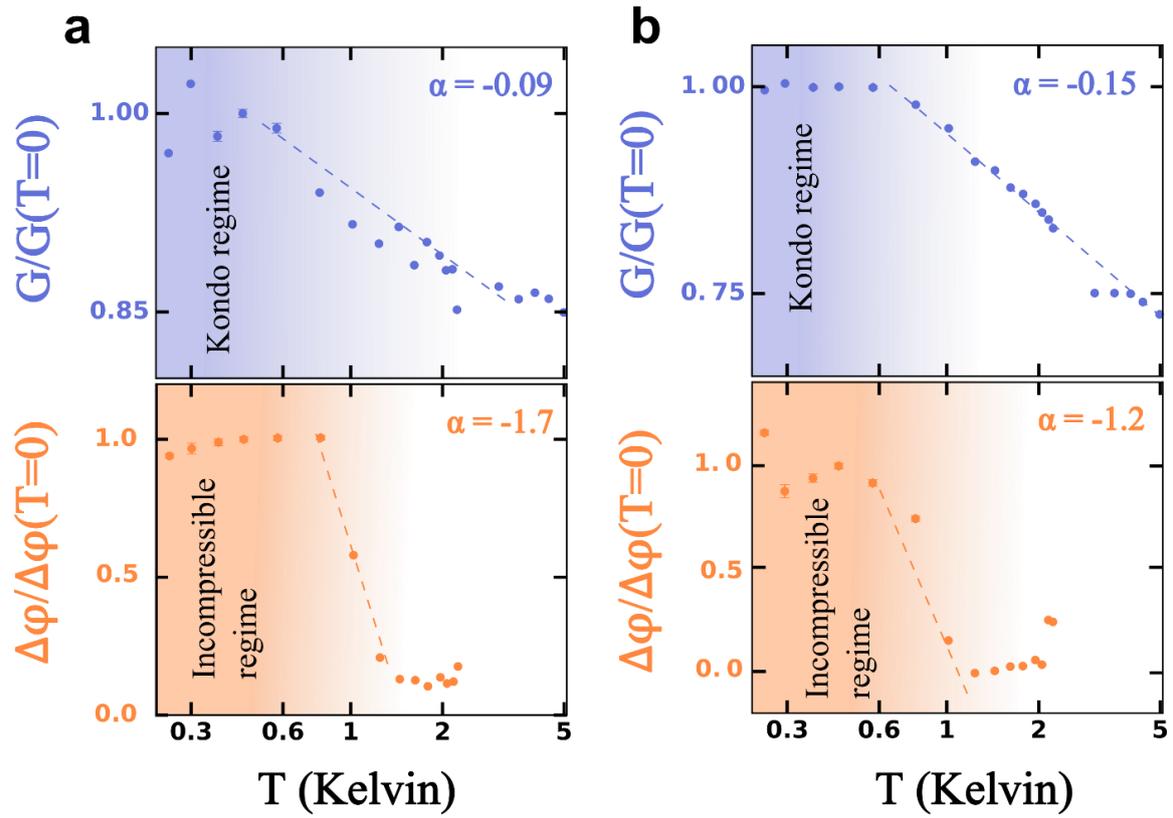

Extended Data Fig. 5
Desjardins et al.

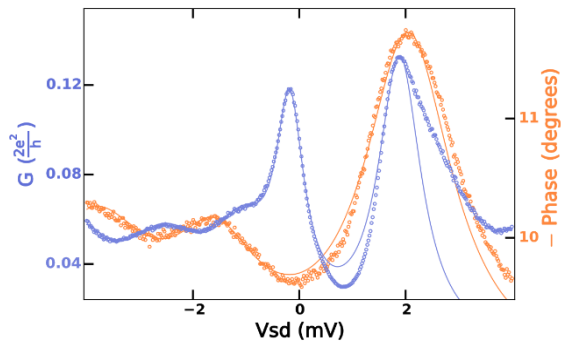
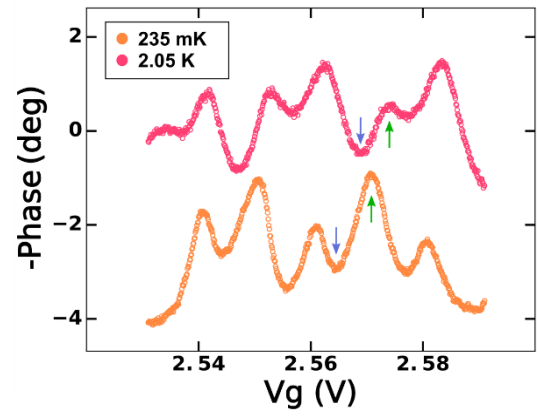

Extended Data Fig. 6
Desjardins et al.

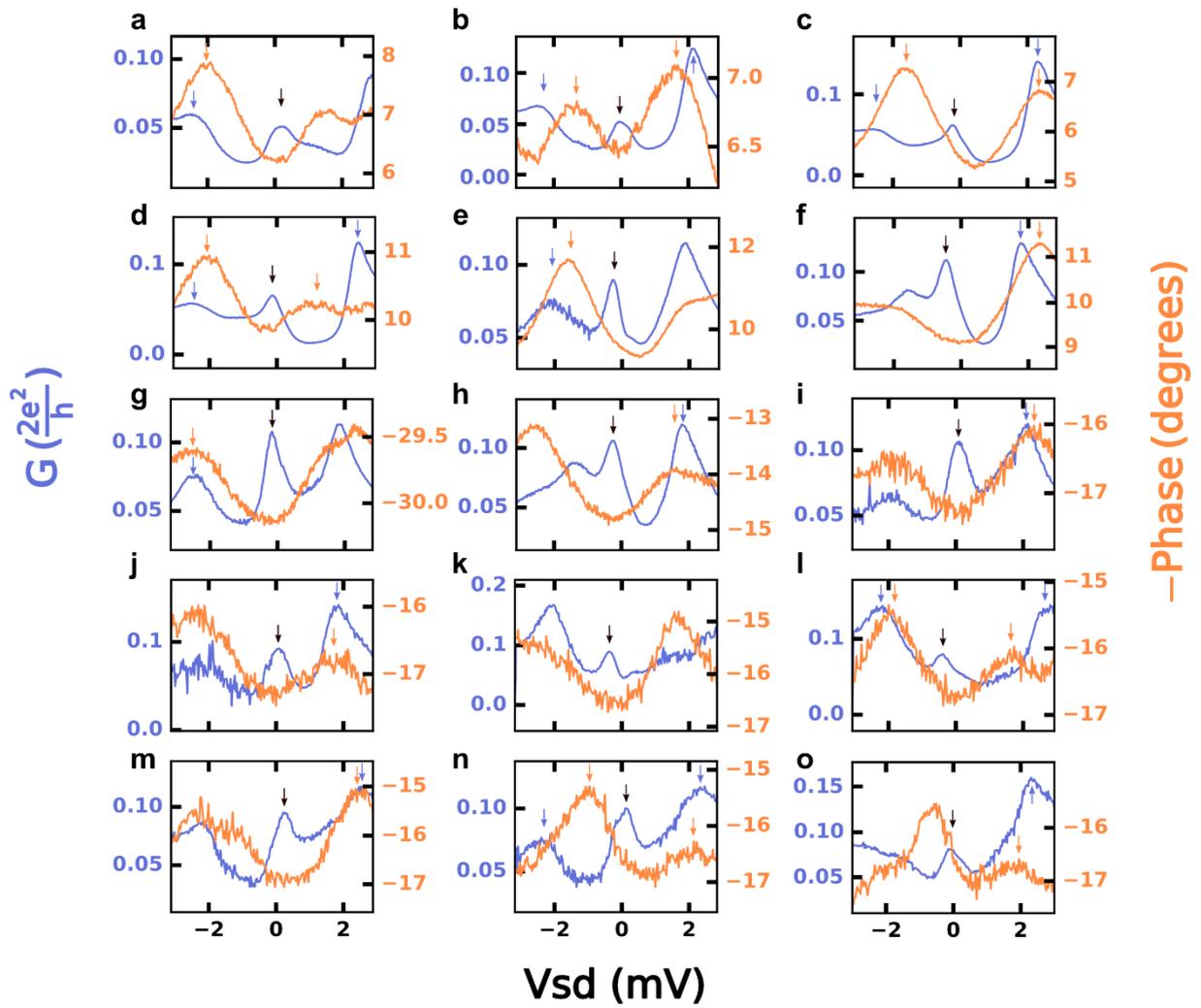

Extended Data Fig. 7
Desjardins et al.

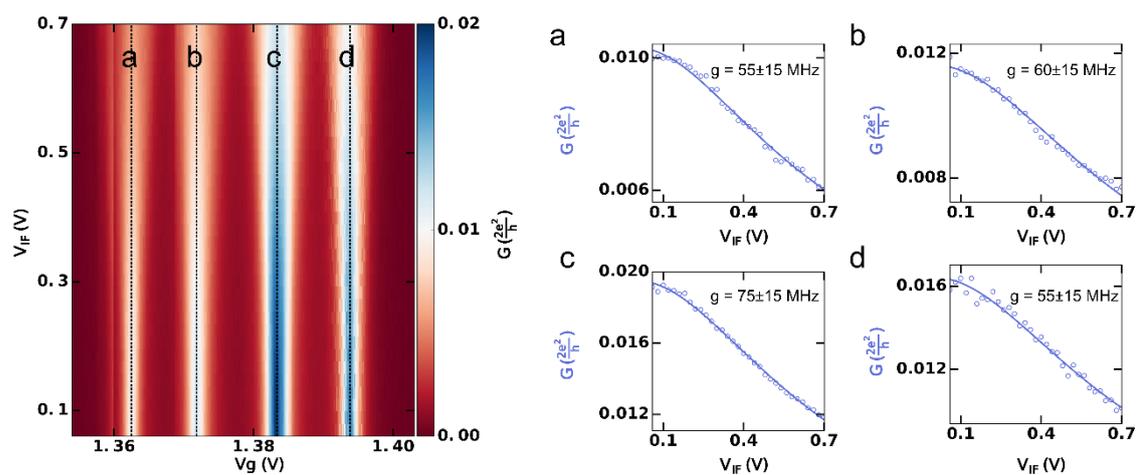

Extended Data Fig. 8
Desjardins et al.